\documentstyle[prl,aps,epsf]{revtex}
\begin{document}
\draft
\bigskip
\title{$\gamma  \gamma\to \pi^0  \pi^0$
as a test for the existence of a light $\sigma$ meson}
\vskip2ex
\author{J.\ L.\ Lucio M., G.\ Moreno, and M.\ Napsuciale}
\address{Instituto de F\'{\i}sica, Universidad de Guanajuato \\
Loma del Bosque \# 103, Lomas del Campestre, 
37150 Le\'on, Guanajuato; M\'exico}

\author{J.\ J.\ Toscano}
\address{Colegio de F\'{\i}sica, FCFM, Benem\'erita Universidad
Aut\'onoma de Puebla \\
Apdo.\ Postal 1152, 72000 Puebla, Pue., M\'exico}
\maketitle

\begin{abstract}
We work out predictions of the Linear Sigma Model for the 
$\gamma \gamma \to\pi^0  \pi^0$ cross section. We consider the sigma width 
which is introduced in  a consistent way  with chiral Ward identities. The 
results of Chiral Perturbation Theory are recovered in
the $m_\sigma \to \infty$ limit. A fit to existing 
experimental results is consistent with a light and broad $\sigma$ meson.

\end{abstract}


\section*{}

Chiral Perturbation Theory (ChPT) is a powerful theoretical 
framework developed to systematically analyze the low energy 
implications of QCD symmetries \cite{pich}. This formalism is 
based on three key ingredients: the chiral symmetry properties 
of QCD, the concept of Effective Field Theory and the fact that, 
below the resonance region $(E < m_R)$, the hadronic spectrum 
only contains an octet of very light --Goldstone bosons--
pseudoscalar particles $(\pi ,K,\eta )$. The resonance region is 
properly defined for each $J^P$ channel, hence the ChPT range of 
validity depends explicitly on the channel being analyzed.

The reaction $\gamma \gamma\to\pi^0 \pi^0$ currently represents 
interesting theoretical \cite{bc,dhl,dh,bgs,penn,bdv,njl} 
and experimental \cite{cball,alde} laboratory for chiral symmetry. 
It has been longly recognized that this reaction is unique 
because its amplitude tests the Chiral Perturbation Theory (ChPT) 
approach at loop level. Indeed, at tree level the amplitude 
vanishes at orders $p^2$ and $p^4$, therefore it must be 
generated at one loop, being necessarily finite and depending 
only on the pion decay constant and meson masses (no arbitrary 
counterterms are possible).

It is important to remark that the agreement between the one loop 
ChPT and $\gamma  \gamma\to\pi^0  \pi^0$ data holds in a small range.
Unfortunately the threshold behavior of the cross section is not 
accounted for by ChPT.
Two potential sources of corrections have been discussed in the literature.
These corrections are:

\begin{enumerate}
\item Rescattering effects required by unitarity, su\-pple\-men\-ted 
with ana\-li\-ti\-ci\-ty as embodied in the dispersive analysis, 
are found to make substantial modifications to the one loop chiral
prediction \cite{dh}.

\item The next to leading order terms in the chiral expansion have 
been evaluated \cite{bgs} and the authors conclude that the improved 
cross section agrees rather well with the data. The enhancement in 
the cross section is mainly due to $\pi \pi$ rescattering and 
the renormalization of the pion decay constant.
\end{enumerate}

\noindent
Another scheme  could be to doubt the threshold datum itself
since this measurement is perhaps the most difficult one. 
Mass resolution, background subtraction, and efficiency corrections 
could affect the cross section measurement. By looking at the data, 
one might tend to average the first two values and consider them as a 
single data point centered at 300~MeV. But even when this is 
done, the trend of the data suggests a faster growth of the cross 
section at threshold than predicted by one-loop ChPT. Under this 
scheme the two loop effects and rescattering effects seem to bring 
theory and experiment into agreement.

An alternative model independent, phenomenological approach was developed by 
Morgan and Pennington \cite{penn}. In that analysis, the experimental $\pi\pi$ 
phase shifts in the $I=0~J=0$ channel are used as an input. The results 
obtained in this approach for the threshold cross section is
very similar to that of two-loop CHPT. 
It still remains however, to understand the dynamics underlying the 
experimental $\pi\pi$ phase shifts.
In particular, the authors  show that the current algebra results for the 
$\pi\pi$ phase shifts yields a poor description of the 
 $\gamma  \gamma\to\pi^0  \pi^0$ cross section near threshold.

Bijnens, Dawson, and Valencia (BDV) \cite{bdv} have studied the Chiral 
Quark model (ChQM) predictions for the reaction 
$\gamma \gamma \to \pi^0  \pi^0$.
The ChQM \cite{mh} describes effective interactions between constituent quark 
fields and the octet of pseudoscalar bosons. Interactions among these fields
are introduced using a non-linear realization of chiral symmetry.
The BDV predictions for the cross section
are in agreement with those of one-loop ChPT and Vector Meson Dominance for 
energies below 400~MeV, thus failing to reproduce the data near threshold.

The process under consideration has been also studied in the 
Nambu-Jona-Lasinio (NJL) model \cite{njl}. The conclusion of this analysis 
is that the enhancement of the cross section near threshold is almost 
accounted for by the tree level corrections of  ${\cal O}(p^6)$ in which a 
sigma is exchanged and that higher order corrections do not affect the low 
energy region.

We face a privileged theoretical and experimental situation which 
deserves a reanalysis of the facts supporting the ChPT approach to 
low energy QCD. Effective Field Theories describe low energy 
physics,
where low is defined with respect to some energy scale $\Lambda$. 
The effective action explicitly contains only those states with mass
$m \ll \Lambda$, while the information of the heavier excitations
is absorbed in the couplings of the resulting low-energy Lagrangian.
One gets in this way
non-renormalizable interactions among the light states, which can be
organized as an expansion in powers of (p/$\Lambda$). In particular,
the behavior of the process we are interested in
($\gamma  \gamma \to \pi^0 \pi^0$), is described in the context of
SU(2)$\times$SU(2) ChPT, with pions as the only degrees of freedom.
However, during the last years evidence \cite{rpp} has accumulated 
for the $f_0(400-1200)$ to form part of the scalar nonet, together 
with $f_0(980)$, $a_0(980)$, and $\kappa(900)$.
If the $f_0(400-1200)$ mass happens to be near the threshold
energy for two pion production, then it must be 
incorporated in the low energy physics description in a  consistent way
with the chiral symmetry of QCD. 

It is our purpose in this letter to work out the predictions for the
behavior
of the $\gamma \gamma\to\pi^0 \pi^0$ cross section
in the low energy region ($\sqrt{s} < 600$ MeV)
within the
context of the Linear Sigma Model (L$\sigma$M). This choice implies
--contrary to
ChPT and ChQM-- the explicit incorporation of the $\sigma$ meson.
Note that, if the sigma width $\Gamma_\sigma$ is considered, 
the requirement of
introducing the sigma meson in a consistent maner with chiral symmetry
is far from trivial. The problem is that the vertices and propagators
are related by Ward identities which have to be respected.
A similar problem ocurrs in the Standard Model where the width of
the gauge bosons have to be included without breaking the gauge
symmetry \cite{grabiel}. In our calculation we modified both the sigma 
propagator
and vertices, ensuring thus the preservation of chiral symmetry.

The need to preserve the symmetry when introducing the sigma meson
shows up clearly in the analysis of $\pi\pi$ scattering.
If the vertices are not properly modified, then the prediction of
the linear sigma model fails to reproduce the derivative coupling
of the pions, typical of the current algebra results associated to the
chiral symmetry. In a separate paper we discuss at length this point
and show that the predictions of the linear sigma model properly
reproduce the experimental $\pi\pi$ phase shifts.

Lorentz covariance and gauge invariance dictate the most general 
form of the amplitude for the 
$\gamma (k_1) + \gamma (k_2) \to\pi^0(k_3)+\pi^0(k_4)$ 
process:

\begin{equation}
{\cal M}={i\alpha\over \pi}{m^2_\pi\over f^2_\pi}\bigl[ A_1(s,t,u)
{P^{\mu\nu}_1 \over s}+ A_2(s,t,u) 
{P^{\mu\nu}_2\over s~k^2_T}\bigr]\epsilon_{1\mu}
\epsilon_{2\nu} \label{eme} ,
\end{equation}

\noindent
where $s,t,u$ are the conventional Mandelstam variables, 
$k_T= \sqrt{(tu-m^4_\pi)/ s}$ denotes the c.m.s. 
transverse (to the incoming photons) momentum of the pion and 
\begin{eqnarray}
P^{\mu\nu}_1&=&k^\mu_2k^\nu_1-{s\over 2}g^{\mu\nu} , \\ \nonumber
P^{\mu\nu}_2&=& m^2_\pi k^\mu_2k^\nu_1+{1\over 2}(tu-m^4_\pi)g^{\mu\nu}
+(m^2_\pi-t)k^\mu_3 k^\nu_1+(m^2_\pi-u)k^\mu_2 k^\nu_3+ s 
k^\mu_3 k^\nu_3 .
\end{eqnarray}

In terms of the $A_i(s,t,u)$ amplitudes, and the scale 
$\Lambda_\chi=4\pi f_\pi$, the cross section is given by:
\begin{equation}
\sigma(\gamma\gamma \to \pi^0\pi^0)=
{\pi\alpha^2\over s^2}\bigl({m_\pi \over 
\Lambda_\chi}\bigr)^4 \int_{t_+}^{t_-} dt \sum_{i=1}^2 |A_i|^2  .
\end{equation}

The SU(2)$\times$SU(2) L$\sigma$M describes the interactions of 
pions and a scalar $\sigma$ meson through the Lagrangian 

\begin{equation}
{\cal L}_{int}= g^\prime \sigma (\sigma^2+\pi^2)+{\lambda\over 4}
(\sigma^2+\pi^2)^2  ,
\end{equation}
\noindent where 
\begin{equation}
g^\prime=\lambda f_\pi,~~~\lambda={m^2_\sigma-m^2_\pi\over 2f^2_\pi}.
\label{couplings}
\end{equation}

In this model, the $\gamma  \gamma \to \pi^0  \pi^0$ amplitude is 
generated at one-loop level by the diagrams shown in Fig.~1(a),
which include pion loops and sigma effects. 
We use a Breit-Wigner type for 
the sigma propagator in order to deal with a possible pole in the $s$ channel.
However, in order to respect chiral Ward identities it is necessary to modify 
also the vertex functions (this point is further elaborated in \cite{lnm}). 
As for 
the present process,  the incorporation of the sigma width in a consistent 
way with 
chiral symmetry is achieved by the replacement 
$m^2_\sigma \to \tilde{m}^2_\sigma \equiv
m^2_\sigma-i \Gamma_\sigma m_\sigma$ in the couplings $\lambda$
and $g^\prime$, given in Eq.~\ref{couplings}.

With these modifications,
the analytical calculation of the amplitudes leads to:
\begin{eqnarray}
A_1(s,t,u)&=&{s-m^2_\pi \over m^2_\pi}\biggl( 1-
{s -m^2_\pi \over s-\tilde{m}^2_\sigma}
\biggr)F(\xi)/2  \label{a1} ,\\ \nonumber
A_2(s,t,u)&=&0 ,
\end{eqnarray}

\noindent where
\begin{equation}
F(\xi)=1+\xi(ln({1+\sqrt{1-4\xi} \over 1-\sqrt{1-4\xi}})-i\pi)^2,~~~
\xi={m^2_\pi\over s}. 
\end{equation}

\noindent
Notice that to lowest order in $s-m^2_\pi$ 
the amplitude given in Eq.\ (\ref{a1}) reproduces the ChPT result
(Compare to Eq.~(37) in Ref.~\cite{dhl}), and that the $\sigma$ meson
induces a next to leading order correction vanishing in the
$m_\sigma \to \infty$ limit idependently of the value of $\Gamma_\sigma$.

The model just described can be extended to include effective quarks.
The linear chiral realization counterpart of the BDV model has been 
discussed in the literature \cite{scad2}. The resulting Linear Chiral
Quark Model (LChQM) contributes with the diagrams shown in Fig.~1(b)
in addition to those in Fig.~1(a).
Elsewhere we report details of the full calculation whose results are 
expressed in terms of the $C_0$ and $D_0$ Veltman-Pasarino 
scalar functions.  The results depend on the quark masses and the 
$\sigma$ meson properties. In our calculation we used
$m_u = m_d = {m_P / 3} $, where $m_P$ stands for the proton mass, 
leaving the mass and width of the $\sigma$ meson as free parameters.
Predictions for the cross section are obtained after the phase space 
integrals are numerically evaluated using the FF program \cite{ff}.

In the LChQM both amplitudes in Eq.\ (\ref{eme}) get non-vanishing
contributions. The loop of pions, the triangle loop of fermions, 
and the box diagrams contribute to the $A_1$ amplitude. $A_2$ only 
receive contributions from the quark box diagrams. We have verified 
that the analytical results possess the appropriate symmetries under
the exchange of identical particles, and we have numerically checked
that our amplitude reproduces the BDV results for
$m_\sigma\to\infty$.

Both L$\sigma$M and LChQM predict a cross section dominated by a
Breit-Wigner peak if it was the case that $\Gamma_\sigma << m_\sigma$.
On the other
hand, a wide Breit-Wigner peak ($ \Gamma_\sigma \sim m_\sigma $) would
appear as a shoulder in the spectrum.
Since we expect that neither L$\sigma$M nor LChQM describe well the
reaction $\gamma  \gamma \to \pi^0  \pi^0$ for energies above
600~MeV, we perform fits to the experimental results considering
only data from the first seven points.
We have phenomenologically incorporated the crossed channel $\rho$
and $\omega$ contributions to the $\gamma  \gamma \to \pi^0  \pi^0$ process
and found that it does not appreciable affect the predictions
of the L$\sigma$M, except for $\sqrt{s} > 600$~MeV, where these
contributions improve the fit. However, at such high energies other
important contributions are expected. Fig.~2 shows a fit of the L$\sigma$M
predictions (no vector mesons included) which yields  a sigma mass and width
of 375 and 365~MeV, respectively.

At threshold our predictions are close to the two-loop ChPT whereas above 
350~MeV the models differ substantially. It is worth remarking that our 
results completelly agree with those obtained in \cite{penn} on the ground of 
general arguments and experimental phase shifts. This is not surprising as 
experimental phase shifts  are well reproduced by the $L\sigma M$ \cite{lnm}.

The large error
bars of existing experimental data makes it impossible to rule out
either L$\sigma$M or ChPT predictions. Higher quality data in the
350--600~MeV region could settle the question, providing thus further
insight on the existence of a low lying scalar meson.
At this point a word of caution is worth. Long ago Mei{\ss}ner \cite{ug}
mentioned that 
the effect of a low lying scalar can be represented by properly summed
pion loops. Comparison of the L$\sigma$M and ChPT at threshold seem
to confirm the results in Ref.~\cite{ug}.
Thus, the
unitarity corrections seen in the two loop effects within the ChPT
scheme are reproduced in the linear realization approach at the one
loop level. It is important to remark, however, that considering the dynamical 
effects of a light scalar meson provides a simple description of data without 
requering  aditional free parameters. 

The incorporation of quarks in the L$\sigma$M may produce a much steeper
growth at threshold than ChPT or the L$\sigma$M, as shown in Fig.~3.
Unfortunatelly, the fit value obtained for the $\sigma$ is 
unreasonable low that we do not pursue the analysis further along
this line.

Summarizing, we point out that the failure of  CHPT ${\cal O}(p^4)$ 
calculations 
in the description of the data for the $\gamma \gamma\to\pi^0 \pi^0$
cross section near threshold  could be due
to the Effective Lagrangian Approximation used in that approach, in which
it is assumed that the only degrees of freedom in the low energy region are 
the pseudoscalar mesons and 
the scalar sigma meson is integrated out from the action. In this letter 
we report the full one loop calculation of the $\gamma \gamma\to\pi^0 \pi^0$
cross section within the sigma model, in which the Chiral Symmetry is 
linearly realized and the sigma meson degrees of freedom are explicitly 
considered. 
The sigma width is introduced in a consistent way with chiral symmetry.
The $L\sigma M$ results 
agree with those obtained in a dispersive 
analysis with experimental $\pi\pi$ phase shifts as input. In the region 
near  
threshold, the $L\sigma M$  also agree with two-loop CHPT calculations, but 
substantially differ above 450 MeV. 
Thus, although present data seems to indicate the existence of a light and 
broad scalar meson, further data on the behavior of the 
$\gamma \gamma\to\pi^0 \pi^0$ cross section could settle the existence 
of such a particle. The DA$\Phi$NE facility will have the 
opportunity to perform a detailed study of the two photon process,
unraveling perhaps a clean signature of the $\sigma$ meson.

Two of us (JLL and MN) wish to acknowledge Prof.\ M.\ Scadron for
usefull discussions.
This work was supported in part by CONACYT of M\'exico under Grant Nos.\
3979P-E9608, 4009P-E9608 and J27604-E.

\bigskip

\vskip2ex
\centerline{
\epsfxsize=350 pt
\epsfbox{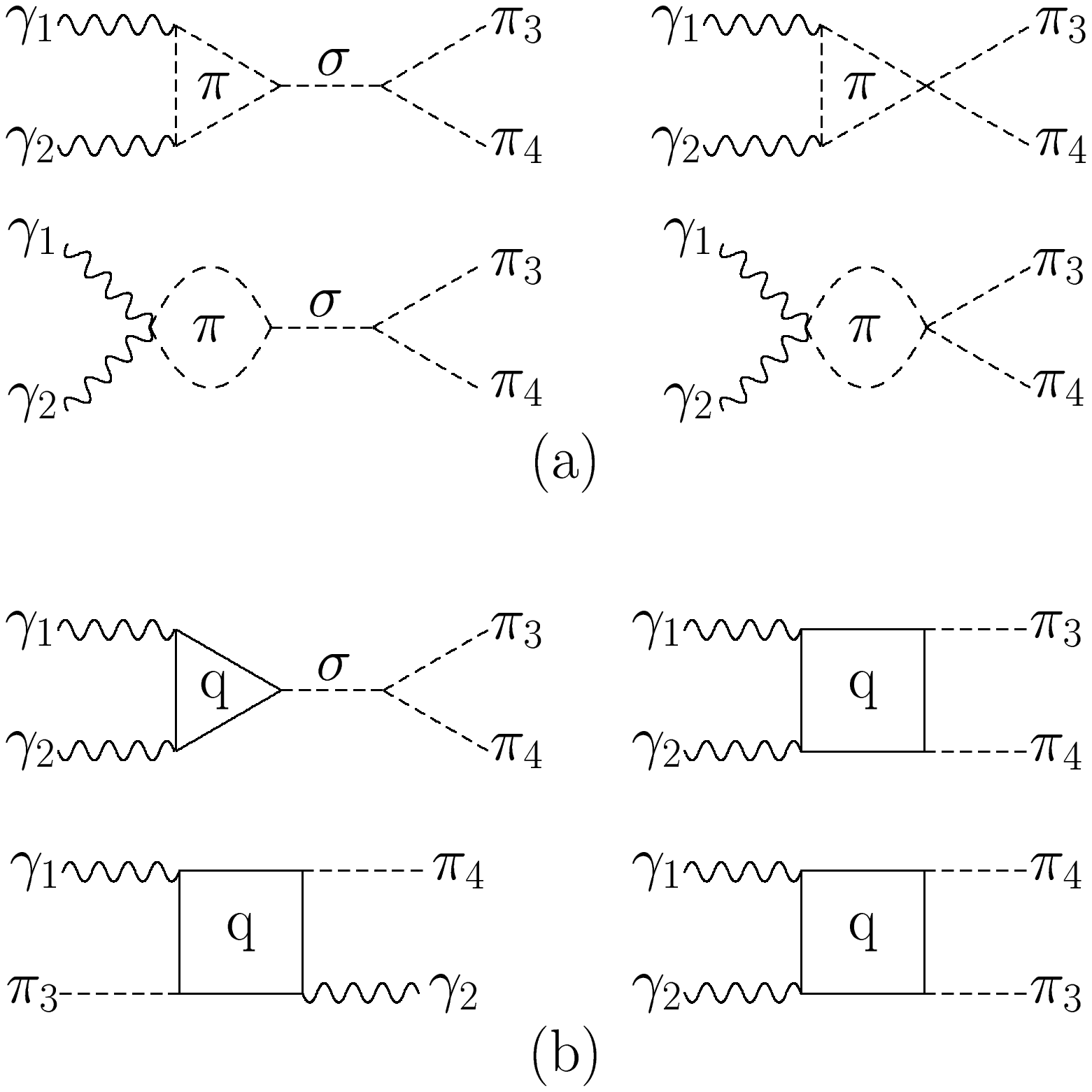}}
\begin{center}
{\small{Fig.1}\\
(a) One-loop level diagrams in the L$\sigma$M 
	and (b) quark contributions of the LChQM for the process
	$\gamma \gamma \to \pi^0 \pi^0$. All diagrams, but the
	seagull type, also contribute with the interchange
	$\gamma_1 \leftrightarrow \gamma_2$. }
\end{center}

\bigskip

\vskip2ex
\centerline{
\epsfxsize=350 pt
\epsfbox{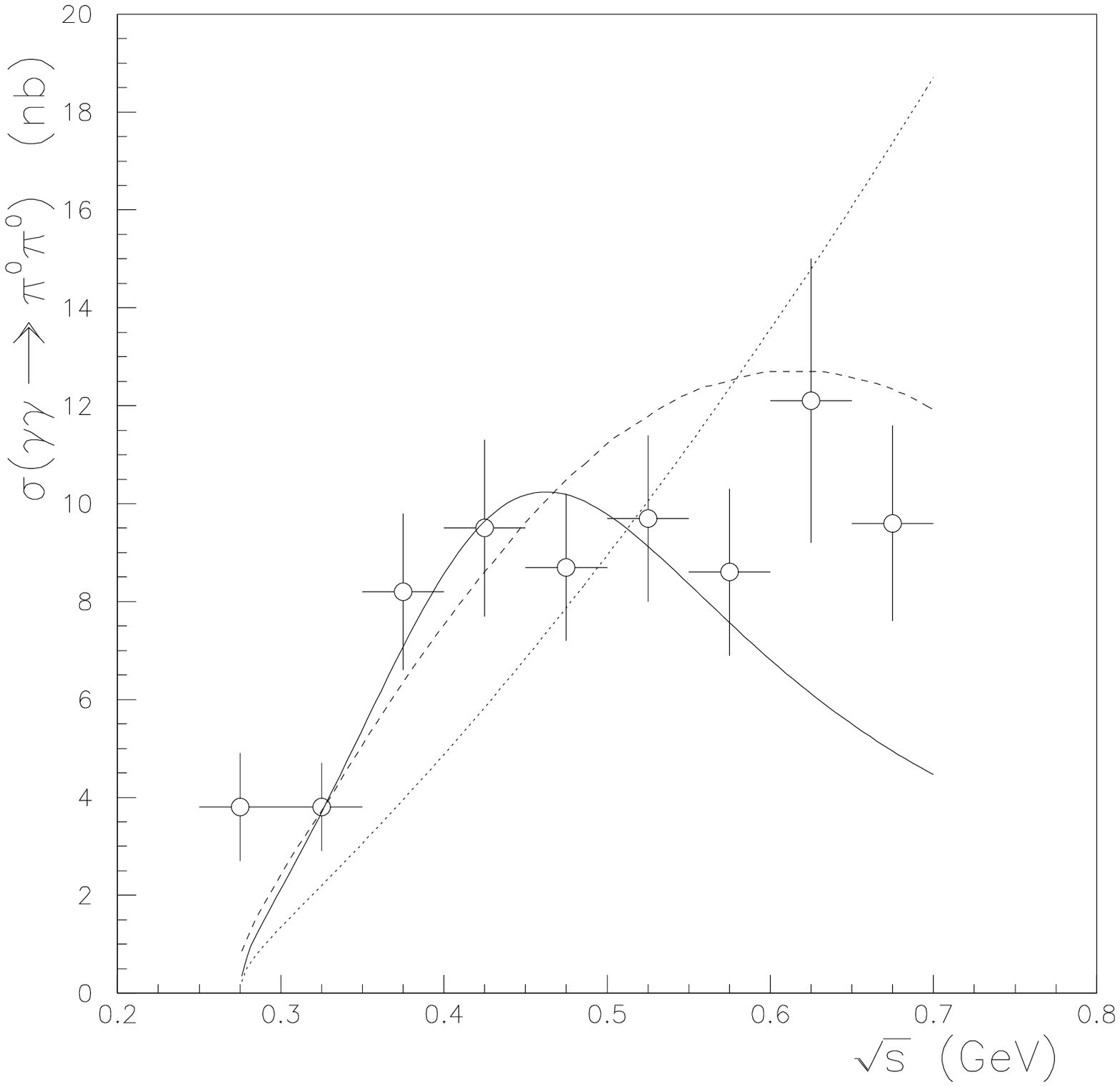}}
\begin{center}
{\small{Fig.2}\\
The solid line is the L$\sigma$M fit
	to the $\gamma \gamma \to \pi^0 \pi^0$ cross section data
	of Marsiske {\it et al.} [8]. The dotted and dashed lines 
	correspond to the one and two loop ChPT predictions, respectively. }
\end{center}

\bigskip

\vskip2ex
\centerline{
\epsfxsize=350 pt
\epsfbox{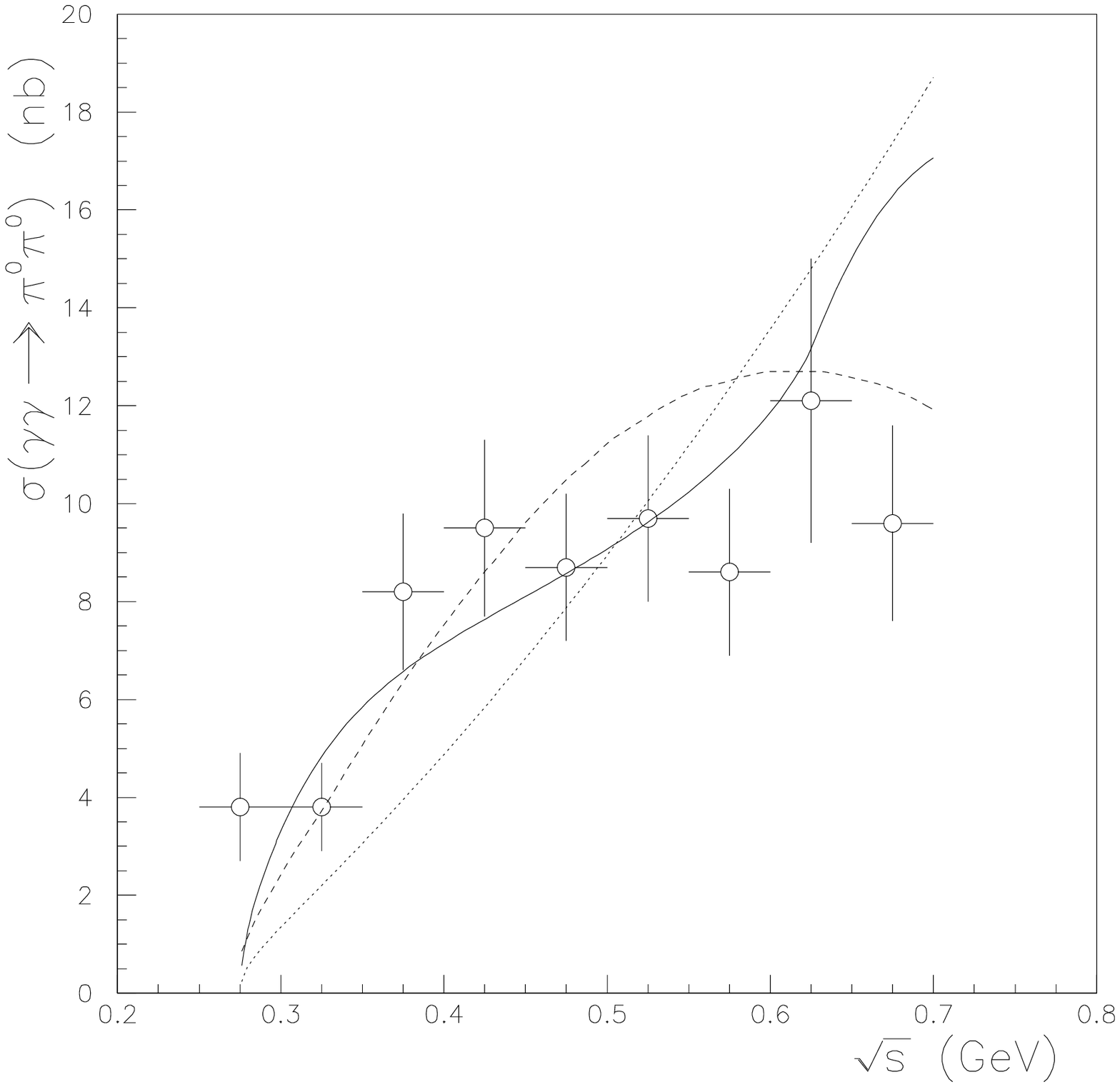}}
\begin{center}
{\small{Fig.3}\\
The solid line is the LChQM fit
        to the $\gamma \gamma \to \pi^0 \pi^0$ cross section data
        of Marsiske {\it et al.} [8]. The dotted and dashed lines correspond
        to the one and two loop ChPT predictions, respectively. }
\end{center}



 
\end{document}